\documentclass[english,aps,manuscript,aps]{revtex4}
\usepackage[T1]{fontenc}
\usepackage[latin9]{inputenc}
\usepackage[letterpaper]{geometry}
\usepackage{float}
\usepackage{amsmath}
\usepackage{graphicx}
\usepackage{amssymb}
\usepackage{esint}

\makeatletter

\providecommand{\tabularnewline}{\\}

\@ifundefined{textcolor}{}
{%
 \definecolor{BLACK}{gray}{0}
 \definecolor{WHITE}{gray}{1}
 \definecolor{RED}{rgb}{1,0,0}
 \definecolor{GREEN}{rgb}{0,1,0}
 \definecolor{BLUE}{rgb}{0,0,1}
 \definecolor{CYAN}{cmyk}{1,0,0,0}
 \definecolor{MAGENTA}{cmyk}{0,1,0,0}
 \definecolor{YELLOW}{cmyk}{0,0,1,0}
 }

\makeatother

\usepackage{babel}

\begin{document}

\title{Maximizing the hyperpolarizability poorly determines the potential}

\author{T. J. Atherton, J. Lesnefsky, G. A. Wiggers and R. G. Petschek}

\address{Case Western Reserve University, 10900 Euclid Avenue, Cleveland,
Ohio, USA 44106}
\begin{abstract}
We have optimized the zero frequency first hyperpolarizability $\beta$
of a one-dimensional piecewise linear potential well containing a
single electron by adjusting the shape of that potential. With increasing
numbers of parameters in the potential, the maximized hyperpolarizability
converges quickly to $0.708951$ of the proven upper bound. The Hessian
of $\beta$ at the maximum has in each case only two large eigenvalues;
the other eigenvalues diminish seemingly exponentially quickly, demonstrating
a very wide range of nearby nearly optimal potentials, and that there
are only two important parameters for optimizing $\beta$. The shape
of the optimized wavefunctions converges with more parameters while
the associated potentials remain substantially different, suggesting
that the ground state wavefunction provides a superior physical description
to the potential for the conditions that optimize the hyperpolarizability.
Prospects for characterizing the two important parameters for near-optimum
potentials are discussed.
\end{abstract}
\maketitle

\section{Introduction}

The non-linear response of a electron to an electric field is an important
fundamental and practical issue in non-linear optics \citep{Dalton:1995p2949,Kuzyk:2000p2905}.
In a notable series of papers, Kuzyk et al. have shown\citep{Kuzyk:2000p2905,Tripathy:2004p2912,Zhou:2006p2911,Zhou:2007p2904}
that all of the measured off-resonant first hyperpolarizabilities
$\beta_{zzz}$ \textemdash{} specifically the second derivative of
the polarization in a given direction $z$ with respect to an electric
field with vanishingly small frequency and wavevector in that direction
\textemdash{} of known molecules are considerably smaller than a theoretical
upper limit $\beta_{max}$ derived from the Thomas-Kuhn sum rules
and the sum-over-states formula\begin{equation}
\beta_{max}=\sqrt[4]{3}\left(\frac{e\hbar}{\sqrt{m}}\right)^{3}\frac{N^{3/2}}{E_{10}^{7/2}}.\label{eq:bmax}\end{equation}
Here $N$ is the number of electrons participating and $E_{10}=E_{1}-E_{0}$
is the energy difference between the ground state and first excited
state. This upper bound \emph{need not} be realizable as it is unclear
that the matrix elements required can be achieved.

One possible strategy to achieve large hyperpolarizabilities is to
optimize the shape of the potential in which the electrons are confined.
Kuzyk et al. have studied a sub-problem of this, namely performing
the optimization for the single-electron case with both one and two
dimensional potentials. They suggest that implementing the qualitative
(modulated) features observed in their optimized potentials might
allow design of molecules with higher hyperpolarizabilities\citep{Zhou:2006p2911,Zhou:2007p2904}.
While these calculations have resulted in a number of potentials with
large hyperpolarizabilities, they do not distinguish which features
of the optimized potential are \emph{required} for high $\beta$ and
which are artifacts of the minimization procedure or chosen parametrization.

In this work we have re-examined the optimization of the first hyperpolarizability
for a single electron moving in a potential in one dimension. We find
convincing evidence that the best possible hyperpolarizability is
substantially smaller than previously published limits: instead we
propose a new lower actual maximum of approximately $0.708951\beta_{max}$.
Moreover, we find a surprisingly broad class of potentials that have
values of $\beta$ essentially indistinguishable from the maximum
and that, in effect, there are only two important constraints on potentials
to have near-optimal $\beta$. Perturbations to an optimized potential
that do not affect its hyperpolarizability include those where the
ground state wavefunction is small as well as rapid changes where
the ground state wavefunction is large. 

To quantify the effect of such perturbations, we present the problem
in the language of optimization theory. Here, one speaks of maximizing
an objective function\textemdash{}in this case the {}``intrinsic''
hyperpolarizability $\beta_{int}=\beta/\beta_{max}$\textemdash{}with
respect to some set of parameters, i.e. those that specify a potential.
Once an optimum value has been found, it is natural to ask how sensitive
that value is to variations in the parameters. This information is
contained in the Hessian matrix, the eigenvalues of which are essentially
the local principal curvatures of the objective function at the maximum\citep{NumericalRecipes,Tarantola}. 

It is quite possible, and will be shown to be the case in the present
work, for these eigenvalues to have very different magnitudes. While
the eigenvalues of the Hessian lack a physical interpretation, their
importance for the numerical optimization procedure is immediate:
large ratios of eigenvalues of the Hessian mean that the maximization
procedure is, in effect, attempting to find the highest point on a
very narrow ridge with a flat top. Generally, optimization programs
will stop when improvement of the function is slow, or the approximate
gradients are small. Alternatively, the program may stop after a specified
number of iterations; the result may be accepted if the value of the
objective function has converged even if the values of the parameters
have not. It is therefore to be expected that the numerical errors
in parameter combinations associated with small eigenvalues of the
Hessian are much larger (by some inverse power of the eigenvalue roughly
between $-1/2$ and $-1$) than those associated with large eigenvalues\citep{Tarantola}.
While these estimated errors, unfortunately, are typically not reported
by such programs, examining them allows one to be confident both that
the actual maximum hyperpolarizability has been found \emph{and }that
the potentials are not inappropriately over-specified. For the present
work, they also provide information on the likely errors in reported
potentials. Since very small eigenvalues indicate a flat maximum,
small errors associated with discretization, numerics, etc may substantially
change the result.

There are multiple possible causes for large ratios of eigenvalues.
It may be that some fundamental feature of the problem is responsible,
or that the representation chosen is deficient in that it admits this
behavior. In the present case, we expect that the hyperpolarizability
depends on the potential primarily where the ground state wave function
is substantial, so parameters that control the shape of the potential
elsewhere are likely to affect the hyperpolarizability negligibly.
Moreover, since the hyperpolarizability has been shown to depend only
on dipole matrix elements and may be derived entirely as a multiple
integral of the ground state wavefunction \citep{Wiggers:2007p2948},
changes in the potential where the ground state wavefunction is large
are also irrelevant if they tend not to alter the wavefunction. For
example, rapid variations in the potential which approximately average
to zero will not significantly affect the hyperpolarizability. For
numerical optimizations like those performed below, it is desirable
to specify the potential in such a way as to avoid including parameters
made redundant by these two effects: the fewer irrelevant parameters
included in such optimizations, the greater the precision that can
be achieved for the remainder.

The number of parameters that characterize the chosen potential is
also important: if too few coefficients are included the problem is
\emph{over}-constrained and the optimal value of $\beta_{int}$ will
be significantly less than the true maximum. Conversely, if too many
coefficients are included the problem might become \emph{under}-constrained.
In such a case many potentials yield comparably optimal values and
interpretation of the potential requires care: important features
may be obscured by irrelevant ones associated with small eigenvalues
of the Hessian. Moreover, the convergence will be slow particularly
for numeric schemes, like those used to date on this problem, that
only evaluate the objective function and not its derivatives.

A known technique\citep{Bretthorst} for investigating such optimization
problems is to increase the number of parameters describing the problem
and examine the eigenvalues and eigenvectors of the Hessian at the
maximum with each additional parameter. To investigate whether the
potential is under-constrained we have chosen the parametrization
as suggested above and varied the number of parameters that characterize
the potential. In each case, we optimized $\beta_{int}$ and calculated
the eigenvalues and eigenvectors of the Hessian at the maximum. This
analysis allows us to state which aspects of the potential are largely
irrelevant to the hyperpolarizability. It also reveals that effectively
only two continuous parameters need to be specified to determine near-optimal
potentials in the present problem.

\section{Model}

There are myriad possible parametrizations for 1D potentials and several
quite simple potentials have hyperpolarizabilities that are a large
fraction of the putative maximum, e.g. the semi-infinite triangular
well and the clipped harmonic oscillator\citep{Zhou:2007p2904}. We
have chosen to represent the potential as a piecewise linear function
that consists of $N+1$ segments\begin{equation}
V(x)=\begin{cases}
A_{0}x+B_{0} & x<x_{0}\\
A_{n}x+B_{n} & x_{n-1}<x<x_{n},\ n\in\{1,2,...,N-1\}\\
A_{N}x+B_{N} & x>x_{N-1}\end{cases}\label{eq:potentialpiecewise}\end{equation}
with positions $x_{n}$ and slopes $A_{n}$ as the parameters. The
$x_{n}$ are strictly ascending and this must be enforced in the optimization.
Since $\beta$ is invariant under both translation of $x$ and the
addition of a constant to $V(x)$, two constants are fixed $x_{0}=0$
and $B_{0}=0$ with no loss of generality. This specificity is desirable:
it decreases the dimensionality of the problem that must be solved
in order to find a given optimum potential, eliminates part of the
null space of the Hessian and so increases the typical size of the
gradient during an optimization. Furthermore, it is substantially
easier to compare optima and verify that the global optimum is unique.
In order to enforce continuity, we take $B_{1}=B_{0}$ and the remaining
constants $B_{n}$ are given by\begin{equation}
B_{n}=\sum_{m=1}^{n-1}(A_{m+1}-A_{m})x_{m},\ n>1.\label{eq:bdefinition}\end{equation}
A remaining degree of freedom irrelevant to the determination of $\beta_{max}$,
the energy scale associated with the potential, is removed by setting
$A_{1}=1$. This does restrict the class of potentials represented
to those in which the leftmost linear segment has negative slope and
the second has positive slope, and so for $N>1$ the effect of choosing
$A_{1}=-1$ was investigated and no improvement in the optimal hyperpolarizability
was found. In order to ensure a bounded wavefunction it is also necessary
to impose $A_{0}<0$ and $A_{N}>0$. Having imposed these constraints,
there remain $2N-1$ free parameters $P_{i}=\{A_{0},A_{2},...,A_{N},x_{1},...,x_{N-1}\}$.
In order to allow the possibility of having an even number of free
parameters, the optimizations were performed in increasing order of
parameter number and for even cases the left hand slope $A_{0}$ was
fixed to the value obtained from optimizing the potential with one
fewer parameter.

The hyperpolarizability for a given potential was calculated by the
following procedure: The Schrödinger equation for one electron in
the $n$-th segment is, in units where $e=1$, $\hbar=1$ and $m_{e}=1$,
\begin{equation}
\left[-\frac{1}{2}\frac{\text{d}^{2}}{\text{d}x^{2}}+(A_{n}+\epsilon)x+B_{n}\right]\psi_{n}=E\psi_{n}\label{eq:schrodinger}\end{equation}
where an electric field of strength $\epsilon$ is also included since
it is desired to calculate $\beta$ at $\epsilon=0$. Its solutions
are the well-known Airy functions\begin{equation}
\psi_{n}(x)=C_{n}\text{Ai}\left[\frac{\sqrt[3]{2}(B_{n}-E+x(A_{n}+\epsilon))}{(A_{n}+\epsilon)^{2/3}}\right]+D_{n}\text{Bi}\left[\frac{\sqrt[3]{2}(B_{n}-E+x(A_{n}+\epsilon))}{(A_{n}+\epsilon)^{2/3}}\right].\label{eq:airyfunctions}\end{equation}
 The $2N+2$ constants $C_{n}$ and $D_{n}$ are determined by imposing
continuity of $\psi(x)$ and its derivative $\psi'(x)$ at each of
the set of points $x_{i}$, yielding $N$ pairs of equations. The
two remaining equations are found by requiring the wavefunction to
vanish as $x\to\pm\infty$, mandating that $D_{0}=D_{N}=0$. For the
case where the number of free parameters $N=2$, the potential is
infinite for $x<x_{0}$ and the two equations at $x=x_{0}$ are replaced
with the usual requirement that the wavefunction vanish there. All
of these equations are linear in the parameters $C_{n}$ and $D_{n}$
and may be represented in matrix form\begin{equation}
W\cdot u=0\label{eq:Wuzero}\end{equation}
where $u$ is the vector of coefficients $\{C_{n},D_{n}\}$ and the
matrix $W$ depends on $E$ and $\epsilon$. Solutions to \eqref{eq:Wuzero}
exist if \begin{equation}
\det W=0,\label{eq:det}\end{equation}
which was readily solved numerically with $\epsilon=0$ to yield an
ordered set of energy levels $\{E_{i}\}$. The hyperpolarizability
$\beta$ was then calculated as follows: The Jacobi formula for the
derivative of the determinant is\begin{equation}
\frac{d}{d\epsilon}\det W=\text{Tr}\left(\text{adj}W\cdot\frac{dW}{d\epsilon}\right)\label{eq:jacobi}\end{equation}
where $\text{adj}W$ is the adjugate of $W$. Since $\det W=0$, for
all $\epsilon,$ $\frac{d}{d\epsilon}\det W=0$. Using the chain rule
\begin{equation}
\frac{dW}{d\epsilon}=\frac{\partial W}{\partial\epsilon}+\frac{\partial W}{\partial E}\frac{dE}{d\epsilon},\label{eq:dWde}\end{equation}
the derivative of the energy with respect to $\epsilon$ may be readily
calculated\begin{equation}
\frac{dE}{d\epsilon}=-\frac{\text{Tr\ensuremath{\left(\text{adj}W\cdot\frac{\partial W}{\partial\epsilon}\right)}}}{\text{Tr\ensuremath{\left(\text{adj}W\cdot\frac{\partial W}{\partial E}\right)}}}.\label{eq:dEde}\end{equation}
Higher derivatives may be obtained by differentiating \eqref{eq:jacobi};
performing this once yields an expression for $\frac{d^{2}E}{d\epsilon^{2}}$\begin{equation}
\frac{d^{2}E}{d\epsilon^{2}}=-\frac{\text{Tr\ensuremath{\left[\frac{d}{d\epsilon}\left(\text{adj}W\right)\cdot\frac{dW}{d\epsilon}+\text{adj}W\cdot W'\right]}}}{\text{Tr\ensuremath{\left(\text{adj}W\cdot\frac{\partial W}{\partial E}\right)}}}\label{eq:d2Ed2e}\end{equation}
where \begin{equation}
W'=\left(\frac{\partial^{2}W}{\partial\epsilon^{2}}+\frac{dE}{d\epsilon}\left[2\frac{\partial^{2}W}{\partial E\partial\epsilon}+\frac{dE}{d\epsilon}\frac{\partial^{2}W}{\partial E^{2}}\right]\right)\label{eq:W'}\end{equation}
and a second time yields $\frac{d^{3}E}{d\epsilon^{3}}$\begin{equation}
\frac{d^{3}E}{d\epsilon^{3}}=-\frac{\text{Tr\ensuremath{\left(\frac{d^{2}}{d\epsilon^{2}}\left(\text{adj}W\right)\cdot\frac{dW}{d\epsilon}+2\frac{d}{d\epsilon}\left(\text{adj}W\right)\cdot\frac{d^{2}W}{d\epsilon^{2}}+\text{adj}W\cdot W''\right)}}}{\text{Tr\ensuremath{\left(\text{adj}W\cdot\frac{\partial W}{\partial E}\right)}}}.\label{eq:deriv3}\end{equation}
where \begin{equation}
W''=\frac{\partial^{3}W}{\partial\epsilon^{3}}+3\frac{d^{2}E}{d\epsilon^{2}}\left(\frac{\partial^{2}W}{\partial E\partial\epsilon}+\frac{dE}{d\epsilon}\frac{\partial^{2}W}{\partial E^{2}}\right)+3\frac{dE}{d\epsilon}\left(\frac{\partial^{3}W}{\partial E\partial\epsilon^{2}}+\frac{dE}{d\epsilon}\frac{\partial^{3}W}{\partial E^{2}\partial\epsilon}\right)+\left(\frac{dE}{d\epsilon}\right)^{3}\frac{\partial^{3}W}{\partial E^{3}}\label{eq:W''}\end{equation}
and then the hyperpolarizability $\beta$ is\begin{equation}
\beta=\frac{1}{2}\left.\frac{d^{3}E}{d\epsilon^{3}}\right|_{E=E_{0},\epsilon=0}\label{eq:betaderiv}\end{equation}
from which the intrinsic hyperpolarizability \begin{equation}
\beta_{int}=\beta/\beta_{max}\label{eq:betaint}\end{equation}
can be calculated using \eqref{eq:bmax} together with the numerically
determined $E_{0}$ and $E_{1}$ from \eqref{eq:det}.

\section{Results and Discussion}

\begin{figure}[H]

\begin{centering}
\includegraphics{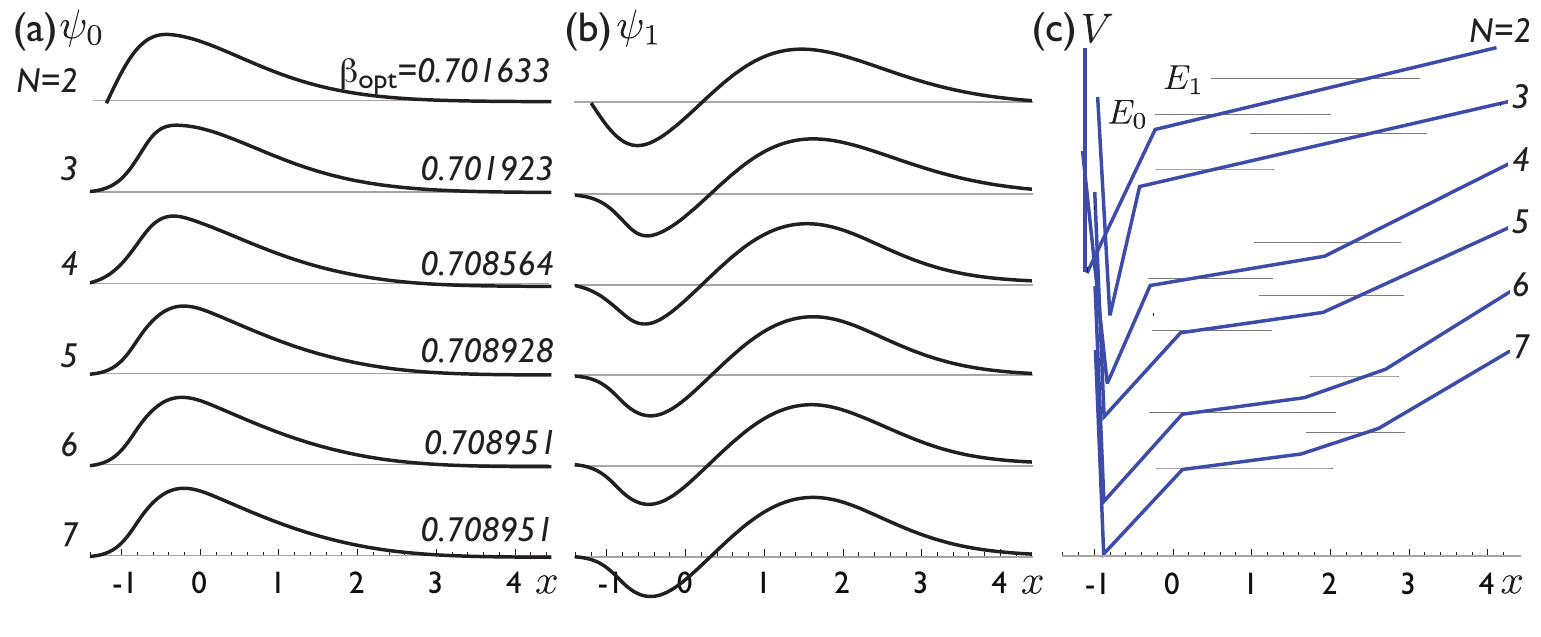}
\par\end{centering}

\caption{\label{fig:Potentials}(Color online) (a) Ground and (b) first excited
normalized wavefunctions corresponding to (c) optimized potentials
for $N=2$ to $N=7$ parameters. Physically irrelevant positions and
energies have been scaled out, consistent with eq. \eqref{eq:scaling}.
The hyperpolarizabilities are noted in (a) and the ground and first
excited energy levels are indicated in (c). }

\end{figure}

A program was written in \emph{Mathematica} 7 using the above formulation
to calculate the intrinsic hyperpolarizability for a potential with
a variable number of parameters and with arbitrary values for those
parameters, expressions for the necessary derivatives in \eqref{eq:deriv3}
being evaluated automatically. The program is provided as supplementary
material to this work. For a given number of parameters ranging from
two to seven, the generic maximization routine \texttt{FindMaximum}
with the Interior Point method and numerically calculated gradients
was used to find locally optimal values of $\beta_{int}$; in each
case the maximization was begun from many different randomly selected
starting points in the parameter space and the best obtained locally
optimized values of $\beta_{int}$ taken to be the global maximum.
We found for larger numbers of parameters that it was necessary to
increase \texttt{WorkingPrecision} to beyond machine precision. A
unique global maximum, to numeric precision, was obtained for each
parametrization independent of the starting point of the optimization.
Care was taken that the optimization had converged to a numerically
accurate result for each parametrization. At each of these globally
optimal values, the Hessian matrix of the objective function\begin{equation}
H_{ij}=\frac{\partial^{2}}{\partial P_{i}\partial P_{j}}\left(\frac{\beta_{int}}{\beta_{max}}\right)\label{eq:Hij}\end{equation}
was calculated numerically with second-order finite differences and
its eigenvalues $h_{j}$ and associated eigenvectors $v_{i}^{j}$
found. For each given number of parameters, the optimized $\beta_{int}$
are displayed in table \ref{tab:table1-2} together with the ratio
of the smallest to largest eigenvalue of the associated Hessian. The
optimized potentials are plotted in fig. \ref{fig:Potentials} together
with the wavefunctions for the ground state and first excited state.
Here the wavefunctions have been normalized and the potentials and
wavefunctions have been shifted and rescaled so that $\left\langle x\right\rangle =0$
and $E_{1}-E_{0}=1$, using \begin{equation}
\bar{x}=(x-\left\langle x\right\rangle )/(E_{1}-E_{0})^{1/2},\ \ \bar{V}(\bar{x},\{P\})=(V(\bar{x},\{P\})-E_{0})/(E_{1}-E_{0})\label{eq:scaling}\end{equation}

This allows us to compare transparently, without the confusion of
irrelevant scaling, the wavefunctions and potentials for different
parametrizations of the potential in order to try to understand which
features are pertinent to optimizing $\beta_{int}$. Note that the
optimal potentials for different numbers of parameters are markedly
different despite the fact that additional parameters generalize the
potential. In contrast, changes in the wavefunctions as the number
of parameters increases are far more subtle. As anticipated, the adjustable
features in the optimized potentials are found to be strictly in the
region where the ground and first excited state wavefunctions are
large, even though this was not enforced in the optimization.

Like previous analyses, we find the very best value of $\beta_{int}$
to be somewhat lower than $1$ which corresponds to a $\beta$ equal
to the theoretical upper limit of \eqref{eq:bmax}. With only four
parameters in the potential, we find a value of $\beta_{int}=0.708764$
comparable to the best previously known, and the highest value so
far obtained $\beta_{int}=0.708951$ with only 6 parameters\textemdash{}though
it is a refinement of only $0.02\%$ over previous results. The addition
of a seventh free parameter yields no further increase in $\beta_{int}$
to within our numerical accuracy. We believe that this value is very
close to the actual maximum achievable for a single electron moving
in one dimension, or more speculatively, for the hyperpolarizability
of a single electron with all components of the electric field along
a single direction as e.g. $\beta_{zzz}$. Formally it is simply a
lower bound for the actual upper bound on $\beta_{int}$. Given that
we have carefully verified that the optimization has converged, it
is clear that there are no potentials similar to these that have $\beta$
significantly higher by more than a fraction of a percent. It remains
possible that there exists a potential with still larger hyperpolarizability
that is not accessible with our parametrization, or which is qualitatively
different from the potentials we have explored. However, to the best
of our knowledge, \emph{all} potentials that have been shown to have
$\beta_{int}$ approaching this value have ground state wavefunctions
similar to those we have found. The wide range of parametrizations
that have been explored, particularly by Kuzyk et al. \citep{Tripathy:2004p2912,Zhou:2006p2911,Zhou:2007p2904},
strongly suggests that qualitatively different but substantially better
potentials do not exist and that the true optimum hyperpolarizability
is close to those presented here. 

\begin{figure}[H]

\begin{centering}
\includegraphics{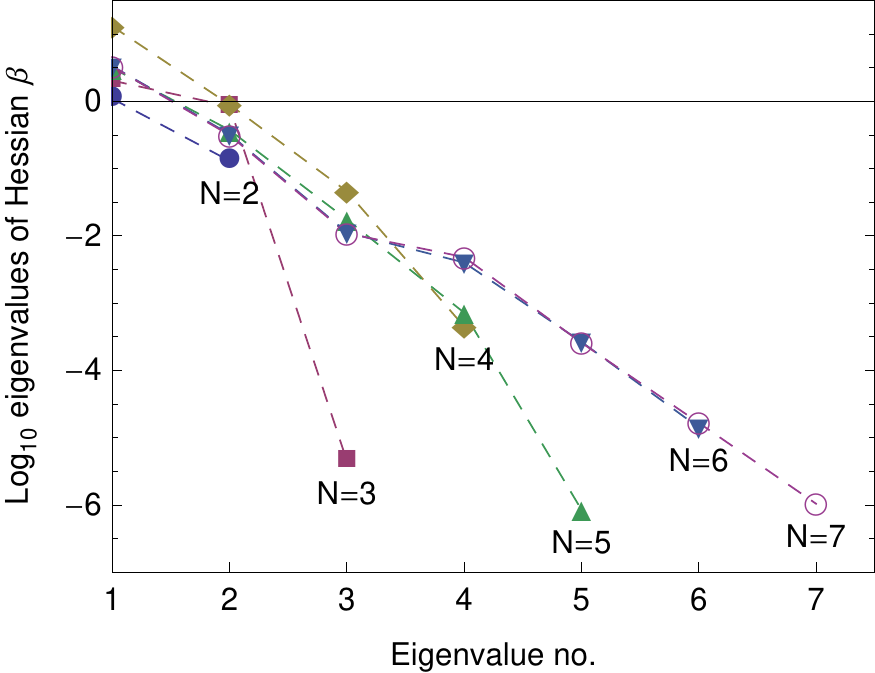}
\par\end{centering}

\caption{\label{fig:Eigenvalues}(Color online) Eigenvalues of the Hessian
of $\beta_{int}$ evaluated at its maximum plotted for different numbers
of free parameters $N$. The dashed lines are intended as a visual
aid only.}

\end{figure}

For our optimized potentials, there is a substantial variation in
the eigenvalues of the associated Hessian as may be seen from table
\ref{tab:table1-2} and fig. \ref{fig:Eigenvalues}. Surprisingly,
it appears that effectively only 2 parameters, those associated with
the two large eigenvalues in the Hessian, need to be adjusted in order
to achieve the maximum hyperpolarizability. There are, therefore,
a very wide range of potentials near to the optimized one that have
hyperpolarizabilities close to the maximum. This is consistent with
the speculation of \citet{Kuzyk:2000p2905} that only some aspects
of the first few wavefunctions\textemdash{}specifically the ratios
of energy differences and dipole matrix elements of the ground and
first two excited states\textemdash{}significantly affect the objective
function: relatively few parameters in the potential are required
to adjust these three low-lying wavefunctions accurately enough to
specify quantities derived from their integral such as dipole matrix
elements. Variations of the potential where these wavefunctions are
small, or rapid variations where they are large, have little effect
on either energies or dipole matrix elements. 

\begin{figure}[H]

\begin{centering}
\includegraphics[scale=0.75]{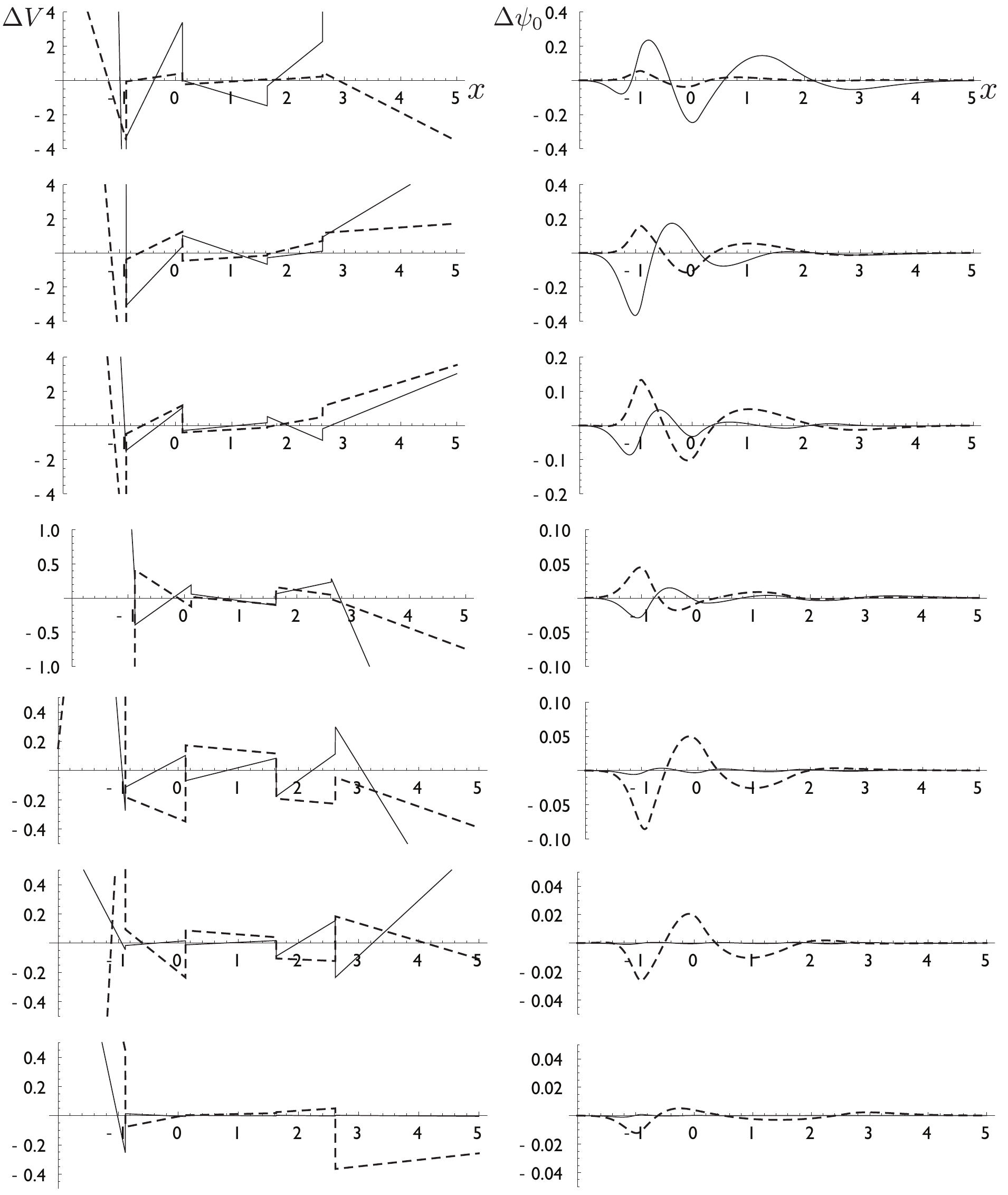}
\par\end{centering}

\caption{\label{fig:DeltaVMeasure}Variations of the potential and ground state
wavefunction associated with eigenvectors of the Hessian matrix of
$\beta_{int}$ evaluated at the maximum; these have been adjusted
so that each variation simultaneously leaves $E_{1}-E_{0}=1$ and
$\left\langle x\right\rangle =0$. The variations are plotted in descending
order of eigenvalue. Each variation is calculated in two measures:
the numerically natural measure (solid lines), and the measure of
the ground state wavefunction (dashed lines).}

\end{figure}

The eigenvectors of the Hessian allow us to determine what aspects
of the optimized potential are pertinent to maximizing the hyperpolarizability.
The variation in the potential $\Delta V^{j}(x)$ in the direction
of the eigenvector associated with the $j$-th largest eigenvalue
of the Hessian $h^{j}$ is \begin{equation}
\Delta V^{j}(x)=\left.\frac{\partial V(x,\{P_{i}+\alpha v_{i}^{j}\})}{\partial\alpha}\right|_{\alpha=0}\label{eq:deltavj}\end{equation}

where $v_{i}^{j}$ is the $i^{th}$ component of the $j$-th eigenvector%
\footnote{Note that, in order to shift and scale the potential and calculate
$\Delta V_{j}$, the derivatives of $\left\langle x\right\rangle $,
$E_{0}$, and $E_{1}$ with respect to $\alpha$ must be calculated,
which requires calculating the derivative of the ground state wavefunction
with respect to $\alpha$. Formulae for these derivatives are given
in \citep{Wiggers:2007p2948}.%
}. A small variation in the potential along this direction $V\rightarrow V+\alpha\Delta V^{j}$
will result in a corresponding change in the hyperpolarizability $\beta_{int}\rightarrow\beta_{int}+\frac{1}{2}h^{j}\alpha^{2}.$
The significance of such a variation is obscured by the fact that
it typically contains some component that serves only to translate
and rescale the energy and length. Since the hyperpolarizability is
not affected at all by these operations, it is desirable to remove
their associated components from the $\Delta V^{j}$. This may be
achieved by rescaling $V$ and $x$ in the right hand side of \eqref{eq:deltavj}
as a function of $\alpha$ to preserve the properties $\left\langle x\right\rangle =0$
and $E_{1}-E_{0}=1$ using \eqref{eq:scaling}. The resulting adjusted
$\Delta V^{j}(x)$ no longer contain irrelevant rescalings or translations
but appear to decrease in apparent size with decreasing eigenvalue
(fig. \ref{fig:DeltaVMeasure}), indicating that they are increasingly
associated with either the irrelevant quantities $\left\langle x\right\rangle $,
$E_{0}$ and $E_{1}-E_{0}$ or with changes where the wavefunctions
are small. Unfortunately, those $\Delta V^{j}(x)$ associated with
large eigenvalues do not seem to converge with increasing numbers
of parameters and so reveal little about which features of the potential
are truly important to the hyperpolarizability.

This suggests our parametrization of the potential could be improved,
and also that a physically more relevant measure of {}``distance''
between potentials is desirable. This measure is implicit in the process
of finding the eigenvalues and eigenvectors of the Hessian; in the
above analysis, it had the form

\begin{equation}
\left\Vert V_{1}-V_{2}\right\Vert ^{2}=\left|P_{1}-P_{2}\right|^{2}\label{eq:apriorimeasure}\end{equation}
where the $P_{i}$ is the vector of parameters that specifies the
$i$-th potential. While this idea of distance\textemdash{}referred
to hereafter as the {}``numerically natural measure''\textemdash{}matters
for the numerical process of finding the maximum, it is largely irrelevant
to the design and synthesis of molecules that have large hyperpolarizabilities.
To explore the possibility that the idea of distance between potentials
matters for the interpretation of what matters for the potentials,
we have constructed more physical measures of the distance between
two potentials $V_{1}$ and $V_{2}$ as follows. First rescale each
potential using \eqref{eq:scaling} to set $\left\langle x\right\rangle =E_{0}=0$
and $E_{1}-E_{0}=1$, yielding scaled potentials $V_{1}^{s},$ $V_{2}^{s}$
for which the quantities irrelevant to the hyperpolarizability have
been adjusted to have standard values. The measure is then

\begin{equation}
\left\Vert V_{1}^{s}-V_{2}^{s}\right\Vert _{k}^{2}=\int_{-\infty}^{\infty}\text{d}x\ \rho_{k}(x)\left|V_{1}^{s}-V_{2}^{s}\right|^{2}\label{eq:measures}\end{equation}
where $\rho_{k}$ is a positive-definite weighting function. There
are a wide variety of plausible possible choices for $\rho_{k}$:
we used $\rho_{0}=\psi_{0}^{2}$ and $\rho_{1,0}=\psi_{0}^{2}+\psi_{1}^{2}$
where $\psi_{0}$ and $\psi_{1}$ are the normalized, scaled wavefunctions
for the optimized potential as these measures serve to remove physically
irrelevant differences between potentials, and are consistent with
\emph{a priori }expectations that only changes in the potential where
low lying wavefunctions are large matter to the hyperpolarizability. 

Eigenvalues of the Hessian were recalculated in each of these measures
and are displayed in table \ref{tab:EigenvaluesMeasure}. Clearly,
the eigenvalues and eigenvectors of the Hessian are quite different
for the different ways of measuring distances; unfortunately, this
is particularly true for the eigenvectors that are associated with
large eigenvalues of the hessian, i.e. those eigenvectors that tell
us which changes to the potential \emph{do} matter to the hyperpolarizability.
Physically this is clear: changes that do not change the hyperpolarizability
can be freely included in changes to eigenvectors with large eigenvalues
if they do not also increase the distance. To examine this more precisely,
consider the generalized eigenvalue problem with different measures
of distances that may be written as \begin{equation}
H_{ki}v_{i}^{j}=h^{j}M_{ki}v_{i}^{j}.\label{eq:measure_eigen}\end{equation}
Here, in a given parametrization, $H_{kj}$ is the hessian of the
objective function ($\beta_{int}$); $M_{kj}$ is the positive definite
hessian of the measure of distance in the same parametrization; the
$v_{i}^{j}$ are the eigenvectors and the $h^{j}$ are the eigenvalues.
As only the eigenvectors and ratios of the eigenvalues are of interest,
$\mathbf{M}$ can be chosen to have unit determinant. Some idea of
the importance of the measure can be obtained by considering two nearly
identical measures and calculating the changes in the eigenvalues
and eigenvectors using perturbation theory in $\delta\mathbf{M}=\mathbf{M}_{1}-\mathbf{M}_{2}$.
Let $h_{\alpha}^{j}$ and $\mathbf{v}_{\alpha}^{j}$ be the eigenvalues
and eigenvectors within measure $\mathbf{M}_{\alpha}$ where eigenvectors
are normalized by $\mathbf{v}_{\alpha}^{k}\mathbf{M}_{\alpha}\mathbf{v}_{\alpha}^{j}=\delta_{jk}$
and $\delta\mathbf{\bar{M}}=\left(\mathbf{M}_{1}\right)^{-1/2}\delta\mathbf{M}\left(\mathbf{M_{1}}\right)^{-1/2}$.
Supposing the eigenvalues $h_{1}^{j}$ and eigenvectors $\mathbf{v}_{1}^{j}$
for the measure $\mathbf{M}_{1}$ are known and that $\delta\bar{\mathbf{M}}_{jk}=\mathbf{v}_{1}^{j}\delta\bar{\mathbf{M}}\mathbf{v}_{1}^{k}$
is small for $j\ne k$, the eigenvalues and eigenvectors $h_{2}^{j}$
and $\mathbf{v}_{2}^{j}$ in the second measure $\mathbf{M}_{2}$
may be found from a routine perturbation calculation: \begin{equation}
h_{2}^{j}=h_{1}^{j}\left(\mathbf{1+v_{1}^{j}}\delta\bar{\mathbf{M}}\mathbf{v_{1}^{j}}\right)+\mathcal{O}\left(\delta\bar{\mathbf{M}}^{2}\right)\label{eq:evals change from measure}\end{equation}
and \begin{equation}
\mathbf{v}_{2}^{j}=\mathbf{v}_{1}^{j}+\sum_{j\ne k}\frac{\mathbf{v}_{1}^{k}h_{1}^{i}\left(\mathbf{v}_{1}^{k}\delta\mathbf{\bar{M}}_{1}\mathbf{v}_{1}^{j}\right)}{h_{1}^{j}-h_{1}^{k}}+\mathcal{O}\left(\delta\bar{\mathbf{M}}^{2}\right).\label{eq:evec chagn from measure}\end{equation}
If $h_{1}^{j}\gg h_{1}^{k}$ any matrix element $\delta\bar{\mathbf{M}}_{jk}$
between these eigenvectors will result in a comparably sized change
in the eigenvector: in other words, the contribution of an eigenvector
that corresponds to a small eigenvalue to an eigenvector with a large
eigenvalue is controlled largely by the measure and not by the nature
of the hyperpolarizability. If, on the other hand, $h_{1}^{k}\ll h_{1}^{j}$,
changes in the measure result in rather little change to the eigenvector.
Unfortunately, the eigenvectors for which $h_{1}^{k}\gg h_{1}^{j}$
are the ones that ought to indicate which features of the optimized
potential most affect the hyperpolarizability. For this reason, the
question of which features of the potential \emph{do} matter to $\beta_{int}$
is much harder to answer than the matter of what \emph{does not}. 

The changes in the scaled potential associated with each of the eigenvectors
of the Hessian are displayed in fig. \ref{fig:DeltaVMeasure} with
the numerically natural measure of eq. \ref{eq:apriorimeasure} and
the measure of eq. \eqref{eq:measures} with $\rho_{0}=\psi_{0}^{2}$;
also plotted are the corresponding changes in the ground and first
excited state wavefunctions. Similar calculations for the other measures
proposed produced similar behavior. Again, the superiority of a wavefunction
description of the problem is clear from the fact that changes in
the wavefunctions for the smaller eigenvalues are substantially smaller
than those in the potential.

The scaled $h^{j}$ and $\Delta V^{j}(x)$ within a physically-motivated
measure unambiguously identify the range of potentials close to the
maximum. The breadth of this range with a variety of measures of distance
suggests that the task of synthesizing near optimum chromophores or,
at least, designing near-optimum potentials, may be easier than has
been previously supposed; on the other hand it appears that while
the parametrizations of the potential used previously have yielded
useful estimates of the upper limit on the hyperpolarizability, they
have also tended to obscure the underlying physics.

To further illustrate this, and to attempt to find a clear exposition
of the important constraints for optimum hyperpolarizabilities, we
calculated and display in table \ref{tab:table1-2} two other physical
quantities: the dipole transition matrix element from the ground to
first excited state $x_{10}=\left\langle 1\left|x\right|0\right\rangle $
and the change in dipole moment between the ground and first excited
state, $\Delta x_{10}=\left\langle 1\left|x\right|1\right\rangle -\left\langle 0\left|x\right|0\right\rangle $.
Both of these have been previously identified\citep{Zhou:2006p2911}
from the sum-over-states formula as being important to optimizing
$\beta$. Importantly, they can also be measured. It is apparent that
both quantities converge with increasing number of parameters, $x_{10}$
to $\approxeq55\%$ of the maximum possible value while less rapidly
$\Delta x_{10}\approx1.26$. Thus there appears to be, consistent
with earlier work, a rather specific optimum value of $x_{10}$ and
a less clear optimum choice for $\Delta x_{10}$. To further understand
this we examined the derivatives of these parameters in the direction
of each of the various eigenvectors of the hessian. These are given
for $N=7$ parameters in table \ref{tab:EigenvaluesMeasure} and clearly
diminish with the associated eigenvalue. The largest eigenvector with
the largest eigenvalue seems quite strongly to constrain a linear
combination of them, which is largely $\Delta x_{10}$ and the next
largest eigenvalue constrains another combination that is largely
$x_{10}$. The fact that $\Delta x_{10}$ is largely controlled by
the largest eigenvalue is unusual given that $x_{10}$ converges more
quickly than$\Delta x_{10}$. Nevertheless these quantities appear
at first sight to offer a superior indication of how far a particular
potential is from the optimum because deviations from the optimum
alter them linearly, while the hyperpolarizability is only altered
quadratically. It may be that combinations of these quantities, together
with some other quantity presently unknown form a better indicator
of the closeness of a potential to optimum, particularly for the second
largest eigenvalue for the derivatives are not terribly large. Moreover,
such ideas must be interpretted with care. It is, for example, physically
clear that constructing a potential with a large barrier between the
states that are important to the hyperpolarizability and a physically
distant state precisely resonant with the first excited state will
result in first and second excited states that are superpositions
of the important excited state and this distant state. This will (of
course) dramatically change $x_{10}$ and $\Delta x_{10}$ without
significant effect on the hyperpolarizability. Nevertheless, for the
single electron problem, and absent motions in special directions,
these parameters, with the specific targets calculated herein seem
reasonable heuristic, experimentally accessible measures of the distance
of a physical system from the optimum.

\section{Conclusions}

We have determined that only 2 parameters are important in optimizing
the zero frequency non-resonant hyperpolarizability of a single electron
in a one-dimensional potential; these are associated with large eigenvalues
of the Hessian of the hyperpolarizability, which are dramatically
larger than the other eigenvalues with any of a number of physically
motivated measures. Even with our careful optimization with few, appropriately
chosen, parameters it is very difficult to distinguish the truly optimal
potential from potentials with hyperpolarizabilities very close to
the maximum. This is clearly a generic problem: there many, many potentials
with virtually identical hyperpolarizabilities that will appear to
most observers quite different from the true optimum potential. It
seems relatively easy, particularly in light of the wide variety of
such potentials obtained by others\citep{Tripathy:2004p2912,Zhou:2006p2911,Zhou:2007p2904},
to design and produce systems whose potentials result in $\beta_{int}$
near the apparent maximum. However, the specific features of potentials
that result in such hyperpolarizabilities are, almost in consequence,
difficult to derive and explain. Our calculation is sufficient to
reveal a wide variety of things that \emph{do not }matter to the optimization,
and to make clear that the optimum is more clearly expressed in terms
of the wavefunction than the potential. While this is a significant
improvement over prior works that take no special steps to understand
which features of the potential are irrelevant, we were still unable
to characterize to our satisfaction in a concise manner what \emph{does}
matter about optimum potentials. We believe that a quite different
calculation\textemdash{}described in part below\textemdash{}would
be required to determine this, though not uniquely as the question
of relevance depends in part on specifying a measure of distance between
potentials and these are motivated by a particular chemical or physical
realization. 

Given present results we are exploring other parametrizations of the
potential: a judicious choice would only include variable features
relevant to the maximization problem. The present work shows both
that the optimized potentials become smoother with additional parameters
and that rapid variations in the potential are irrelevant; this implies
that the actual optimum potential is analytic. A promising approach
is to expand the potential in a series of analytic functions such
as Hermite polynomials and to maximize the hyperpolarizability with
respect to the coefficients. Such a parametrization has some attractive
features: If the optimal potential is analytic, the expansion should
converge quickly with the higher order polynomials clearly corresponding
to rapid variations in the potential that are known to be irrelevant.
The following observations: that additional parameters make the potentials
smoother; the eigenvalues of the hessian decrease rapidly; and that
the eigenvectors somewhat resemble hermite polynomials with two added
to the order; all suggest that the best potential is analytic as these
characteristics would be expected for analytic potentials. Second,
the numerically natural measure of distance between potentials induced
by the weighting function of the Hermite polynomials is a physically
reasonable one, since the weighting function itself resembles the
electron's probability density in the ground state. We therefore believe
this parametrization ought to reveal more rigorous lower bounds on
the maximum possible hyperpolarizability and a clearer description
of the optimal potential. It is also expected, for numerically feasible
optimizations, to appreciably decrease errors in the eigenvalues and
eigenvectors of the hessian from purely numerical sources such as
discretization and truncation. This, in turn, may make physical interpretation
of these quantities, and specifically the parameters that most matter
to the hyperpolarizability appreciably easier.

The consequences of our main results for the design of new non-linear
chromophores are less obvious. Our clearest result is that numerically
optimized potentials, such as those obtained in this work and elsewhere\citep{Zhou:2006p2911,Zhou:2007p2904},
need very careful interpretation prior to use as design paradigms:
both the parametrization and the optimization method are likely to
result in appreciable {}``noise'' that is without physical consequence.
It seems remarkably difficult from our (or other) work to specify
the range of nearly optimal potentials \emph{per se} in any way that
will matter to practical design. The $N=2$ and $N=7$ potentials
of figure 1 differ by only slightly more than 1\% in the optimum hyperpolarizability
but appear physically dramatically different. 

We have also demonstrated convincingly that the true upper bound on
the hyperpolarizability of a single electron is appreciably below
the bound derived from the dipole matrix element sum rules. With $n$
electrons, the maximum possible hyperpolarizability is predicted to
be $n^{3/2}$ times that for a single electron. In fact this can be
achieved in a system in which there are strong attractive interactions
between the electrons that consequently move as a single quasi-particle
in a potential similar that calculated above\citep{Watkins:2011p3121}.
This mechanism is likely only to be realizable in exotic systems due
to the Coulomb repulsion of the electrons. A more practical lower
limit on the bound is $n$ times the single electron maximum hyperpolarizability,
which would correspond to the situation where the electrons are confined
to non-interacting wells. Including repulsion, we speculate that the
sum-rule bound increasingly overestimates the true bound with increasing
numbers of electrons: two non-interacting electrons with opposite
spins and Fermi statistics have twice, not $2^{3/2}$ times the upper
bound on the susceptibility derived from the sum rules, while $n$
non-interacting bosons would have only $n$ rather than $n^{3/2}$
times the one particle bound. The fact that the bound for non-interacting
electrons is smaller than that for attractive electrons suggests that
it may also be a bound on electrons that repel each-other. Hence,
it would appear that extant materials fall less spectacularly short
of actual bounds than is presently claimed, although there is likely
still substantial improvement possible in the hyperpolarizabilities. 

Most importantly, it appears that surprisingly few parameters of the
potential need to be adjusted to achieve close to optimum hyperpolarizability.
Thus, while we \emph{do} believe that significantly better chromophores
can still in principle be made, it seems likely that careful, multidimensional
optimization of potentials is \emph{not} required to do this. Rather,
only a few salient features need to be understood and adjusted appropriately.
Beyond the obvious and well known ideas that the potential should
be substantially, but not overly, asymmetric, and that the chromophore
should absorb light well to the first excited state but only to some
fraction of the sum-rule constraint, these features are as yet poorly
understood.

\newpage{}%
\begin{table}
\begin{tabular}{ccccc}
\hline 
No. of Parameters & Optimal $\beta_{int}$ & $h_{min}/h_{max}$ & $x_{10}$ & $\Delta x_{10}$\tabularnewline
\hline
1 & 0.659525 & $0$ &  & \tabularnewline
2 & 0.701633 & $1.19\times10^{-1}$ &  & \tabularnewline
3 & 0.701923 & $2.25\times10^{-6}$ & 0.554604 & 1.354420\tabularnewline
4 & 0.708764 & $3.45\times10^{-5}$ & 0.557740 & 1.268598\tabularnewline
5 & 0.708928 & $2.37\times10^{-7}$ & 0.557752 & 1.269431\tabularnewline
6 & 0.708836 & $4.24\times10^{-6}$ & 0.557762 & 1.269147\tabularnewline
7 & 0.708951 & $3.17\times10^{-7}$ & 0.557763 & 1.269151\tabularnewline
\hline
\end{tabular}

\caption{\label{tab:table1-2}For different numbers of parameters in the potential:
(from the left) Optimized intrinsic hyperpolarizabilities for , together
with the ratio of eigenvalues of the Hessian matrix of the intrinsic
hyperpolarizability. }

\end{table}

\begin{table}
\begin{tabular}{ccccc}
\hline 
\multicolumn{3}{c}{Eigenvalues of hessian} & \multicolumn{2}{c}{Derivatives of dipole matrix elements}\tabularnewline
Numerically natural Measure & $\rho_{0}$ & $\rho_{10}$ & $x_{10}'$ & $\Delta x_{10}'$\tabularnewline
\hline
$-3.28$ & $-22.2$ & $-8.10\times10^{-1}$ & $-3.21\times10^{-1}$ & $2.15$\tabularnewline
$-3.14\times10^{-1}$ & $-4.70$ & $-1.76\times10^{-1}$ & $-7.07\times10^{-2}$ & $5.35\times10^{-3}$\tabularnewline
$-1.09\times10^{-2}$ & $-1.15$ & $-8.19\times10^{-2}$ & $-2.05\times10^{-3}$ & $4.96\times10^{-3}$\tabularnewline
$-4.83\times10^{-3}$ & $-2.47\times10^{-1}$ & $-3.16\times10^{-2}$ & $-1.47\times10^{-3}$ & $1.13\times10^{-2}$\tabularnewline
$-2.58\times10^{-4}$ & $-8.03\times10^{-2}$ & $-5.54\times10^{-3}$ & $-9.80\times10^{-5}$ & $3.91\times10^{-4}$\tabularnewline
$-1.66\times10^{-5}$ & $-4.65\times10^{-3}$ & $-3.03\times10^{-3}$ & $-1.23\times10^{-5}$ & $-7.71\times10^{-6}$\tabularnewline
$-1.04\times10^{-6}$ & $-3.12\times10^{-4}$ & $-2.02\times10^{-4}$ & $3.44\times10^{-7}$ & $-4.85\times10^{-7}$\tabularnewline
\hline
\end{tabular}

\caption{\label{tab:EigenvaluesMeasure}Eigenvalues of Hessian for $N=7$ parameters
in each of three measures: (from the left) the numerically natural
measure induced by the parametrization; the measure $\rho_{0}=\left|\psi_{0}\right|^{2}$
and $\rho_{10}=\left|\psi_{0}\right|^{2}+\left|\psi_{1}\right|^{2}$.
Derivatives of the dipole matrix in the direction of the eigenvector
associated with each eigenvalue are also displayed.}

\end{table}

\end{document}